\documentclass[amsmath,amssymb,
showpacs,twocolumn,
superscriptaddress,
prl]{revtex4-1}
\usepackage{graphicx,bm,color,subfigure}
\usepackage[T1]{fontenc}
\usepackage{mathrsfs}
\usepackage{bm}
\usepackage{amsmath}
\usepackage{amssymb}
\usepackage{graphicx}
\usepackage{esint}
\usepackage{multirow}
\usepackage{float}
\usepackage{array}
\usepackage{makecell}
\usepackage{harpoon}
\usepackage{booktabs}
\usepackage{gensymb}
\usepackage{simplewick}
\usepackage{subfigure}
\usepackage{soul}


\begin{document}
\title{Making Chiral Topological Superconductivities from Non-topological Superconductivities Through the Twist}
\author{Yu-Bo Liu}
\thanks{These two authors contributed equally to this work.}
\affiliation{School of Physics, Beijing Institute of Technology, Beijing 100081, China}
\author{Jing Zhou}
\thanks{These two authors contributed equally to this work.}
\affiliation{Department of Science, Chongqing University of Posts and Telecommunications,
	Chongqing 400065, China}
\author{Yongyou Zhang}
	\affiliation{School of Physics, Beijing Institute of Technology, Beijing 100081, China}
\author{Wei-Qiang Chen}
	\affiliation{Department of Physics and Shenzhen Institute for Quantum Science and Engineering, Southern University of Science and Technology, Shenzhen 518055, China}
	\affiliation{Shenzhen Key Laboratory of Advanced Quantum Functional Materials and Devices, Southern University of Science and Technology, Shenzhen 518055, China}
\author{Fan Yang}
\email{yangfan_blg@bit.edu.cn}
\affiliation{School of Physics, Beijing Institute of Technology, Beijing 100081, China}
\date{\today}
\begin{abstract}
In this paper, we propose a general scheme to realize chiral TSCs through the ``{\it twistronics}''. Suppose we have a $D_n$-symmetric monolayer superconductor, which carries non-topological SC with pairing angular momentum $L=n/2$. Here we propose that we can obtain chiral TSC with the same $L$, by stacking two such monolayers with the largest twist angle $\pi/n$, forming a Moireless quasi-crystal (QC) structure, dubbed as the twist-bilayer QC (TB-QC) here. The chiral TSC in the TB-QC is driven by the interlay Josephson coupling between the pairing order parameters of the two layers. An argument based on the universal Ginzburg-Landau theory is provided to understand this proposal. One known example which fits our proposal is the $d+id$- chiral TSC in the 45\degree-twisted bilayer cuprates. Here, based on the microscopic framework developed previously to treat with the electron-electron interactions in the TB-QC, we demonstrate the application of our proposal to a new example, i.e. the $f+if$- chiral TSC obtained by twisting two properly-doped honeycomb-Hubbard-model monolayers by the angle 30\degree.  This example is related to the newly synthesized 30\degree- twisted bilayer graphene.
\end{abstract}

\maketitle
\section{Introduction}

The origin and physical properties of topological quantum states is an important research area, which has captured great research interests in recent years \cite{Kane2010,Shoucheng2011}. Among these topological states, the chiral topological superconductivities (TSCs) \cite{Read2000} are particularly interesting since such quantum states are characterized by nonzero Chern numbers and resultant Majorana zero modes in the vortex core or on the boundary \cite{Volovik1999}, which carry non-Abelian statistics \cite{Ivanov2001} and can be used in the design of fault-tolerant quantum computation \cite{Kitaev2003, Nayak2008}. The chiral TSCs on a 2D lattice are usually generated by the nontrivial rotational symmetries of the lattice. On an n-fold rotation symmetric 2D lattice, the pairing symmetries can be classified according to the irreducible representations (IRRPs) of the point group \cite{Sigrist1991}. In the special case when the leading pairing symmetry belongs to the 2D IRRP, the two degenerate pairing gap functions would usually be $1:\pm i$ mixed to lower the free energy below the superconducting $T_c$ \cite{Cheng2010}, leading to the chiral TSC. For example, the square lattice possessing $D_4$ symmetry can host the $p+ip$-wave TSC with pairing angular momentum $L=1$ \cite{Yao2015, Meng2015}, while the triangle-, the honeycomb- and the Kagome- lattices which possess the $D_6$ symmetry can host $p+ip$- or $d+id$- wave TSC carrying pairing angular momentum $L=1$ \cite{Ma2014,Chen2015,Zhang2015} or $L=2$ \cite{Doniach2007, Gonzalez2008, Honerkamp2008, Pathak2010, McChesney2010, Nandkishore2012, Wang2012,Kiesel2012,Honerkamp2014,Liufeng2013, ZhangLiDa2015}.

The material realization of the chiral TSC has long been a challenging problem. The Sr$_2$RuO$_4$ used to be a promising candidate of the $p+ip$ chiral TSC \cite{Maeno2003} on the square lattice, but now more and more experimental evidences don't support such a point of view. The quarter-doped graphene \cite{Doniach2007, Gonzalez2008, Honerkamp2008, Pathak2010, McChesney2010, Nandkishore2012, Wang2012,Kiesel2012,Honerkamp2014} and the properly-doped magic-angle twisted bilayer graphene \cite{Xu2018, YangFan2018,WuFeng20181,Isobe2018,You2019,Gonzalez2019} were proposed to host the $d+id$ chiral TSC, but unambiguous experiment evidences are still lack. It's interesting to ask the question whether we can design chiral TSC from existing materials through some engineering approach. Fortunately, the recently emergent ``{\it twistronics}'' gives us the hope. Recently, it was proposed \cite{Yu_Bo2023,cuprates_QC, JPHu2018, cuprates_QC2,cuprates_QC3} that through twisting two cuprate monolayers by the angles near 45\degree \cite{Zhu2021, Zhao2021}, one can obtain the $d+id$ chiral TSC through the interlayer Josephson coupling (IJC). Here the cuprates monolayer with $D_4$-symmetric square-lattice structure is already superconducting, which hosts the $d$-wave SC carrying pairing angular momentum $L=2$. But the $d$-wave pairing on the square lattice belongs to the 1D $B$- IRRP with non-degenerate real gap function, which is non topological. However, when two such non-topological superconducting monolayes are stacked with the proper twist angle, the chiral TSC can be achieved! Note that the special angle 45\degree~here is actually the largest possible twist angle between the two four-fold symmetric monolayer. It's interesting to generalize such a constructive proposal to more lattices with different symmetries.

In this paper, we propose a general scheme to realize chiral TSCs through the ``{\it twistronics}''. Suppose we have a $D_n$-symmetric monolayer superconductor, which carries non-topological SC with pairing angular momentum $L=n/2$, e.g. the $d$-wave SC for square lattice or $f$-wave one for the triangle, honeycomb or Kagome lattice. Here we propose that we can obtain chiral TSC with the same pairing angular momentum, i.e. the $d+id$- or $f+if$- TSC, by stacking two such monolayers with the largest twist angle $\pi/n$ between them, forming a Moireless quasi-crystal (QC) structure, dubbed as the twist-bilayer QC (TB-QC) here. The chiral TSC in the TB-QC is driven by the interlay Josephson coupling (IJC) between the pairing order parameters (ODPs) of the two layers. An argument based on the universal Ginzburg-Landau (G-L) theory is provided to understand this proposal. Then based on the microscopic framework developed previously to treat with the electron-electron (e-e) interactions in the TB-QC, we demonstrate the application of our proposal with a new example different from the known one, i.e. the $f+if$- chiral TSC obtained by twisting two properly-doped honeycomb-Hubbard-model monolayers by the angle 30\degree.  This example is related to the newly synthesized 30\degree- twisted bilayer graphene \cite{Ahn2018,Yao2018,Pezzini2020,Yan2019,Deng2020}.

The remaining part of the paper is organized as follow. In Sec. II, we provide the G-L theory based analysis, which considers what pairing state would be obtained in a TB-QC when each of its monolayer hosts a pairing state with the largest pairing angular momentum for the lattice. In Sec.III, we provide an example to demonstrate the conclusion achieved in Sec. II, i.e. a TB-QC with each of its monolayer described by a Hubbard model on the honeycomb lattice. In Sec. IV, a conclusion is arrived after some discussions.

\section{The G-L theory}
We start from the classification of pairing symmetries on a 2D lattice according to the IRRPs of its $D_{n}$ ($n$ is even hereafter) point group  \cite{Yu_Bo2023}. The $D_{n}$ point group has four 1D IRRPs and $\left(\frac{n}{2}-1\right)$ 2D ones (labeled as $E_L$ ($L\in\left[1,\frac{n}{2}-1\right]$)). For each 2D IRRP $E_L$, the two degenerate basis gap functions would generally be mixed as $1:\pm i$ to lower the free energy. The resultant gap function $\Delta^{(\pm)}_L(\mathbf{k})$ transform as $\Delta^{(\pm)}_L(\mathbf{k})\to e^{\mp iL\Delta\phi}\Delta^{(\pm)}_L(\mathbf{k})$ under a $\Delta\phi=2\pi/n$ rotation, corresponding to a TSC with pairing angular momentum $L\le \frac{n}{2}-1$, and pairing chirality ``$+$'' or ``$-$''. The four 1D IRRPs correspond to the non-topological $A_{1,2}$ pairing symmetry with $L=0$ and $B_{1,2}$ one with $L=\frac{n}{2}$. Here the label $L$ denotes the pairing angular momentum. Clearly, for a $D_n$-symmetric lattice, the largest $L$ is $\frac{n}{2}$, and the pairing with this $L$ is non topological. While for $n=4$ the largest $L=2$ corresponds to the $d$-wave pairing, for $n=6$ the largest $L=3$ corresponds to the $f$-wave one.

\begin{figure}[htbp]
\centering
\includegraphics[width=0.5\textwidth]{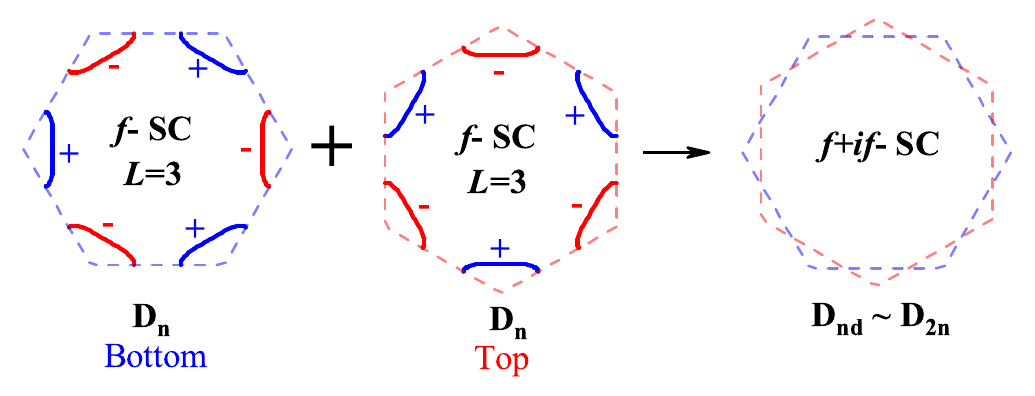}
\caption{(Color online) Schematic illustration of the SC induced by IJC in a TB-QC formed by two $D_n$-symmetric monolayers carrying SC with pairing angular momentum $L=\frac{n}{2}$. We take $n=6$ for example. The color denotes the gap sign on the Fermi surfaces.}\label{schematic_shown}
\end{figure}

Let's take two $D_n$-symmetric monolayers, and stack them by the twist angle $\frac{\pi}{n}$ to form a TB-QC, as shown in Fig.~\ref{schematic_shown} for $n=6$. Obviously, the point group is $D_{nd}$, isomorphic to $D_{2n}$. There is an additional symmetry generator in the TB-QC absent its monolayer, i.e. the $C^{1}_{2n}$ rotation accompanied by a succeeding layer exchange, renamed as $\tilde{C}^{1}_{2n}$ here.

Suppose that driven by some pairing mechanism, the monolayer $\mu=\text{t/b}$ (top/bottom) can host a pairing state with the largest pairing angular momentum $L=n/2$. The pairing gap function in the $\mu$ layer is
\begin{equation}\label{gap_function}
\Delta^{(\mu)}(\mathbf{k})=\psi_{\mu}\Gamma^{(\mu)}(\mathbf{k}).
\end{equation}
Here $\Gamma^{(\mu)}(\mathbf{k})$ is the normalized real form factor, and $\psi_{\mu}$ is the ``complex pairing amplitude''. Prominently, the $\Gamma^{(\mu)}(\mathbf{k})$ for $L=n/2$ changes sign with every $C_n^1$ rotation, due to the following relation
\begin{equation}
e^{iL\Delta\phi}=e^{i\pi}=-1.
\end{equation}
The geometry shown in Fig.~\ref{schematic_shown} dictates
\begin{equation}\label{relation_gap_updn}
\Gamma^{(\text{b})}(\mathbf{k})=\hat{P}_{\frac{\pi}{n}}\Gamma^{(\text{t})}(\mathbf{k}),~~ \hat{P}_{\frac{2\pi}{n}}\Gamma^{(\mu)}(\mathbf{k})=-\Gamma^{(\mu)}(\mathbf{k}).
\end{equation}
Here $\hat{P}_{\phi}$ indicates the rotation by the angle $\phi$. As the interlayer coupling in the TB-QC is weak \cite{Moon2019, Park2019, Yuan2020,Yu_Bo2023}, we can only consider the dominant intralayer pairing. However, the two intralayer pairing ODPs can couple through the IJC describing the combined hopping of a Cooper pair between the two layers \cite{cuprates_QC, JPHu2018, cuprates_QC2,cuprates_QC3}. We shall investigate the ground state induced by this IJC, based on a G-L theory analysis.

The symmetry-allowed free energy $F$ as function of $\psi_{\text{t/b}}$ can be decomposed into the monolayers $F_0(\left|\psi_{\mu}\right|^2)$ term and the IJC $F_J$ term as \cite{cuprates_QC, JPHu2018, cuprates_QC2,cuprates_QC3}
\begin{eqnarray}\label{G_L_F}
F\left(\psi_{\text{t}},\psi_{\text{b}}\right)&=&F_0(\left|\psi_{\text{t}}\right|^2)+F_0(\left|\psi_{\text{b}}\right|^2)+F_{J}\left(\psi_{\text{t}},\psi_{\text{b}}\right),
\end{eqnarray}
Up to the first-order IJC, the $F_J$ term should take the following U(1)-gauge symmetry allowed form,
\begin{eqnarray}\label{G_L_F}
F^{(1)}_{J}\left(\psi_{\text{t}},\psi_{\text{b}}\right)&=&-A\left(e^{i\theta}\psi_{\text{t}}\psi_{\text{b}}^*+c.c\right).
\end{eqnarray}
The invariance of the free energy $F$ under the time-reversal (TR) operation: $\psi_{\text{t/b}}\to \psi^{*}_{\text{t/b}}$ dictates $\theta=0$, leading to
\begin{eqnarray}\label{Josephson}
F^{(1)}_{J}\left(\psi_{\text{t}},\psi_{\text{b}}\right)&=&-A\left(\psi_{\text{t}}\psi_{\text{b}}^*+c.c\right).
\end{eqnarray}

Note that the TB-QC possesses an additional symmetry absent in each of its monolayer, i.e. the $\tilde{C}^{1}_{2n}$ symmetry. Under the $\tilde{C}^{1}_{2n}$ operation, the gap function on the $\mu$ layer changes from $\Delta^{(\mu)}(\mathbf{k})=\psi_\mu\Gamma^{\left(\mu\right)}(\mathbf{k})$ to $\tilde{\Delta}^{(\mu)}(\mathbf{k})=\psi_{\bar{\mu}}\hat{P}_{\frac{\pi}{n}}\Gamma^{\left(\bar\mu\right)}(\mathbf{k})$ which, under Eq. (\ref{relation_gap_updn}), can be rewritten as $\tilde{\psi}_\mu\Gamma^{\left(\mu\right)}(\mathbf{k})$ with
\begin{equation}\label{ODP_change}
\tilde{\psi_\text{b}}=\psi_\text{t}, ~~~~~~~~~~\tilde{\psi_\text{t}}=-\psi_\text{b}.
\end{equation}
The invariance of $F$ under $\tilde{C}^{1}_{2n}$ requires $A=0$, indicating that the first-order IJC should vanish in the TB-QC.

Therefore, the following U(1)-gauge and TR symmetries allowed second-order IJC should be considered,
\begin{eqnarray}\label{G_L_F_2}
F_J\left(\psi_{\text{t}},\psi_{\text{b}}\right)=A_0\left(\psi_{\text{t}}^2\psi_{\text{b}}^{2*}+{\rm c.c.}\right)+O\left(\psi^6\right).
\end{eqnarray}
Eq. (\ref{G_L_F_2}) is minimized at $\psi_b=\pm i \psi_t$ for $A_0>0$ or $\psi_b=\pm \psi_t$ for $A_0<0$. The form case is usually energetically favored as the $1:i$ mixing manner of the ODPs from the two layers leads to complex gap function which is fully gapped. In such a case, under the $\tilde{C}^{1}_{2n}$, the pairing gap function $\left(\psi_b,\psi_t\right)$ would be changed to $\left(\tilde{\psi_b},\tilde{\psi_t}\right)=\left(\psi_t,-\psi_b\right)=\mp i\left(\psi_b,\psi_t\right)=e^{\mp i\frac{2\pi L}{2n}}$ with $L=\frac{n}{2}$, suggesting a chiral TSC belonging to the $E_{n/2}$ IRRP of the $D_{nd}$ or $D_{2n}$ point group.

{\bf To summarize, taking a TB-QC formed by two $D_n$-symmetric monolayer, when each monolayer hosts a non-topological pairing state with pairing angular momentum $L=n/2$, the TB-QC would most probably host a chiral TSC with the same $L$, driven by the IJC.}

While the sign of the above coefficient $A_0$ of the second-order IJC cannot be determined by the G-L theory itself, it should be determined by the microscopic calculations. Previous microscopic calculations favor $A_0>0$ for the 45\degree-twisted bilayer cuprates \cite{Yu_Bo2023,cuprates_QC}. In the next section, we shall do a microscopic calculation to determine the sign of the coefficient $A_0$ for a TB-QC made from two monolayers described by the honeycomb lattice Hubbard model which hosts the $f$-wave SC. It would be shown that the $f+if$ chiral TSC would be obtained for this TB-QC.

\section{Microscopic calculations}
\begin{figure}[htbp]
	\centering
	\includegraphics[width=0.5\textwidth]{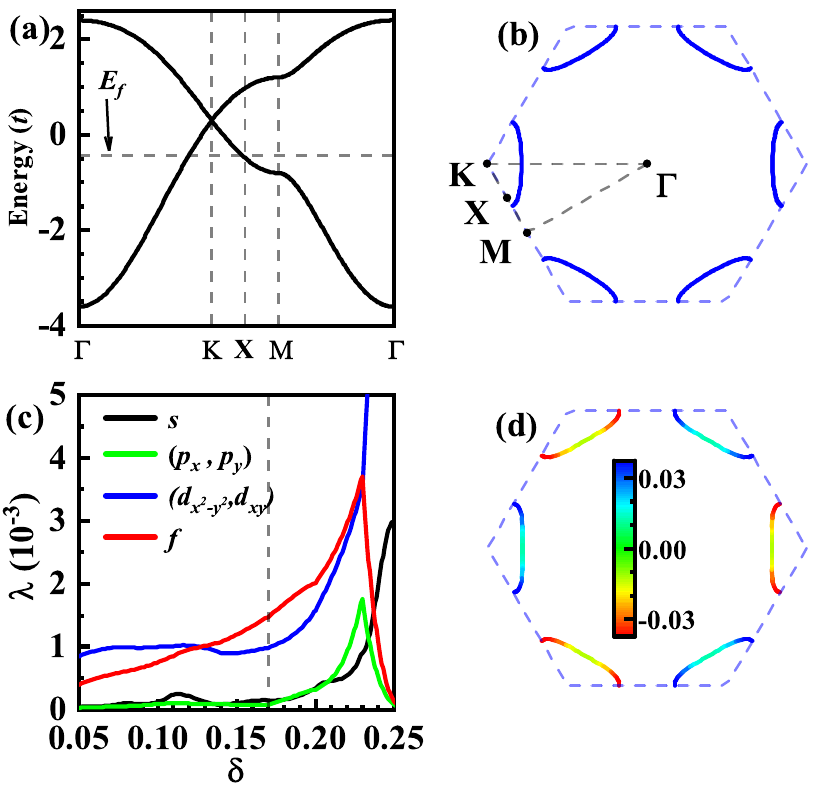}
	\caption{(Color online) Results for the monolayer honeycomb lattice. (a) Band structure along the high-symmetry lines. (b) FS for $\delta = 0.17$ hole doping. (c) The $\lambda \sim \delta$ relations for various leading pairing symmetries at $U=0.3t$. The dashed line marks the doping level $\delta=0.17$. (d) The distribution of the gap function for the $f$-wave SC on the FS.}\label{mono}
\end{figure}
In this section, we provide an example to demonstrate the universal conclusion obtained on the above, which is different from the known one, i.e. the 45\degree twisted bilayer cuprates. Here we choose the honeycomb lattice as an example for $n=6$, and study the Hubbard model. We shall first study the pairing symmetry on a monolayer by the random-phase-approximation (RPA) approach. It will be seen that in the phase diagram obtained, there exists a doping regime in which the $f$-wave SC is the leading pairing symmetry. Then we study the 30\degree-twisted bilayer system of the lattice. Through the microscopic framework developed previously \cite{Yu_Bo2023}, we shall obtain the band structure of this TB-QC. Using the RPA approach, we shall further study the pairing state of the bilayer for a typical doping point at which the monolayer hosts the $f$-wave pairing symmetry. Consequently, our results unambiguously yield the $f+if$ chiral TSC state for this TB-QC.

\subsection{The $f$-wave pairing for the monolayer}

Let's consider the following monolayer honeycomb-lattice Hubbard model with nearest-neighbor (NN) and next-nearest-neighbor (NNN) hopping terms,
\begin{eqnarray}\label{mono_hamiltonian}
H=&-&t\sum_{\left\langle\mathbf{i,j}\right\rangle\sigma}c^{\dagger}_{\mathbf{i}\sigma}c_{\mathbf{j}\sigma}+h.c.-t^{\prime}\sum_{\left\langle\left\langle\mathbf{i,j}\right\rangle\right\rangle\sigma}c^{\dagger}_{\mathbf{i}\sigma}c_{\mathbf{j}\sigma}
+h.c.\nonumber\\&+&U\sum_{\mathbf{i}}n_{\mathbf{i}\uparrow}n_{\mathbf{i}\downarrow}
\end{eqnarray}
The corresponding hopping parameters $t,t^{\prime}$ satisfy $t^{\prime}=0.1t$. For the Hubbard interaction parameter $U$, we have taken a typical $U=0.3t$ friendly for the RPA calculations. The result for larger $U$ below the critical one for the spin density wave instability is qualitatively the same.

The band structure of the model in the absence of $U$ is shown in Fig.~\ref{mono}(a) along the high-symmetry lines, and the Fermi surface (FS) for a typical hole doping $\delta=0.17$ is shown in Fig.~\ref{mono}(b). Obviously, Fig.~\ref{mono}(b) illustrates a sextuple symmetric FS including two hole pockets centering around the K-points. Using the standard multi-orbital RPA approach \cite{RPA1,RPA2,RPA3,RPA4}, we obtain the largest pairing eigenvalue $\lambda$ and the corresponding pairing eigenvector for each pairing symmetry for a given doping level. Here $\lambda$ is related to the $T_c$ via $T_c\propto e^{-\frac{1}{\lambda}}$, and therefore the pairing symmetry with the largest $\lambda$ is the leading one. The relative pairing gap function of the leading pairing symmetry is given by the normalized pairing eigenvector.

The doping $\delta$ dependence of the largest pairing eigenvalue $\lambda$ for various leading pairing symmetries including the non-degenerate $s$-wave, the degenerate $\left(p_x,p_y\right)$-wave, the degenerate $\left(d_{x^2-y^2},d_{xy}\right)$-wave and the non-degenerate $f$-wave are shown in Fig.~\ref{mono}(c). Clearly, the $f$-wave pairing occupies the doping regime $\delta\in(0.1,0.17)$ to be the leading pairing symmetry, which is qualitatively consistent with Ref. \cite{Kiesel2012}. The distribution of the gap function for the obtained $f$-wave pairing is shown in Fig.~\ref{mono}(d) on the FS. This gap function changes sign with every 60\degree- rotation, and it hosts six nodes along the Brillouin-Zone (BZ) diagonal direction.

\subsection{The $f+if$-wave pairing for the TB-QC}

Then let's stack two monolayers described by Eq.~(\ref{mono_hamiltonian}) by the twist angle 30\degree, to form a TB-QC. The total Hamiltonian now reads
\begin{equation}\label{interlayer_hopping}
H=-\sum_{\mathbf{ij}\sigma}t_{\mathbf{ij}}c^{\dagger}_{\mathbf{i}\sigma}c_{\mathbf{j}\sigma}+U\sum_{\mathbf{i}}n_{\mathbf{i}\uparrow}n_{\mathbf{i}\downarrow}
\end{equation}
Here the index $\mathbf{i}$ labels all the sites belonging to both layers, and $t_{\mathbf{ij}}$ represents the hopping integral between the sites $\mathbf{i}$ and $\mathbf{j}$ which can locate in either the same or different layers. For $\mathbf{i}$ and $\mathbf{j}$ locating within the same layer, the formula of $t_{\mathbf{ij}}$ has been given by Eq.~(\ref{mono_hamiltonian}).  The formula of $t_{\mathbf{ij}}$ for the interlayer hopping is given as \cite{Moon2019}
\begin{equation}\label{tij}
t_{\mathbf{ij}}=t_{\mathbf{ij}\pi}\left [1-\left(\frac{\mathbf{R}_{\mathbf{ij}}\cdot\mathbf{e}_{\mathbf{z}}}{R}\right)^{2}\right]+
t_{\mathbf{ij}\sigma}\left(\frac{\mathbf{R}_{\mathbf{ij}}\cdot\mathbf{e}_{\mathbf{z}}}{R}\right)^{2},
\end{equation}
with
\begin{equation}
t_{\mathbf{ij}\pi}=t_{\pi}e^{-\left(R_{\mathbf{ij}}-a\right)/r_{0}},\quad
t_{\mathbf{ij}\sigma}=t_{\sigma}e^{-(R_{\mathbf{ij}}-d)/r_{0}}.\nonumber
\end{equation}
Here, $R_\mathbf{ij}$ is the length of the 3D vector $\mathbf{R}_\mathbf{ij}$ pointing from $\mathbf{i}$ to $\mathbf{j}$, and $\mathbf{e}_{\mathbf{z}}$ is the unit vector perpendicular to the layer. The parameters $a\approx0.142$ nm, $d\approx0.335$ nm, $r_{0}\approx0.0453$ nm, $t_{\pi}=t\approx 2.7$ eV and $t_\sigma\approx -0.48$ eV denote the lattice constant, interlayer spacing, normalization distance, in-plane hoping and vertical hoping, respectively. For these parameters, we have adopted the corresponding parameters for the 30\degree-twisted bilayer graphene \cite{Moon2019}.

As the QC structure doesn't possess translation symmetry, the traditional band-structure theory cannot apply to the electronic structure of this material. However, due to the large twist angle, the interlayer hybridization is weak, and the perturbational-band theory \cite{Ahn2018, Yao2018, Moon2013,Koshino2015, Yu_Bo2023} is suitable to treat with the electronic structure. To involve the e-e interaction, we adopted the following revised perturbational-band theory \cite{Yu_Bo2023} developed previously by some of the authors of this paper.

Concretely, we decompose the tight-binding part of the Hamiltonian into the zeroth-order intralayer hopping term $H_0$ and perturbational interlayer tunneling term $H^{\prime}$. We first diagonalize $H_0$ in the $\mathbf{k}$-space to obtain its eigen state $\left|\mathbf{k}\alpha^{(\text{t/b})}\right\rangle$ and eigen energy $\varepsilon_{\mathbf{k}\alpha}^{\text{t/b}}$. The $H^{\prime}$ can be written as a hybridization form between top-layer state $\left|\mathbf{k}\alpha^{(\text{t})}\right\rangle$ and bottom-layer state $\left|\mathbf{q}\beta^{(\text{b})}\right\rangle$. Consequently, for a given $\left|\mathbf{k}\alpha^{(\text{t})}\right\rangle$ state from the top layer, only a few isolate $\left|\mathbf{q}\beta^{(\text{b})}\right\rangle$ state from the bottom layer can couple with it, justifying the perturbational treatment. Gathering all the $\left|\mathbf{q}\beta^{(\text{b})}\right\rangle$ related to $\left|\mathbf{k}\alpha^{(\text{t})}\right\rangle$, we can calculate the perturbation-corrected eigen state and eigen energy brought about by the $H^{\prime}$ term numerically, which are labeled as $|\widetilde{\mathbf{k}\alpha^{(\text{t})}}\rangle$ and $\tilde{\varepsilon}^{\text{t}}_{\mathbf{k}\alpha}$. Similarly, we get $|\widetilde{\mathbf{q}\beta^{(\text{b})}}\rangle$ and $\tilde{\varepsilon}^{\text{b}\beta}_{\mathbf{q}}$. We have checked that different $|\widetilde{\mathbf{k}\alpha^{(\mu)}}\rangle$ thus obtained are almost mutually orthogonal, qualifying $\{|\widetilde{\mathbf{k}\alpha^{(\mu)}}\rangle\}$ as a good set of single-particle bases to facilitate the succeeding studies involving e-e interaction.

\begin{figure}[htbp]
	\centering
	\includegraphics[width=0.5\textwidth]{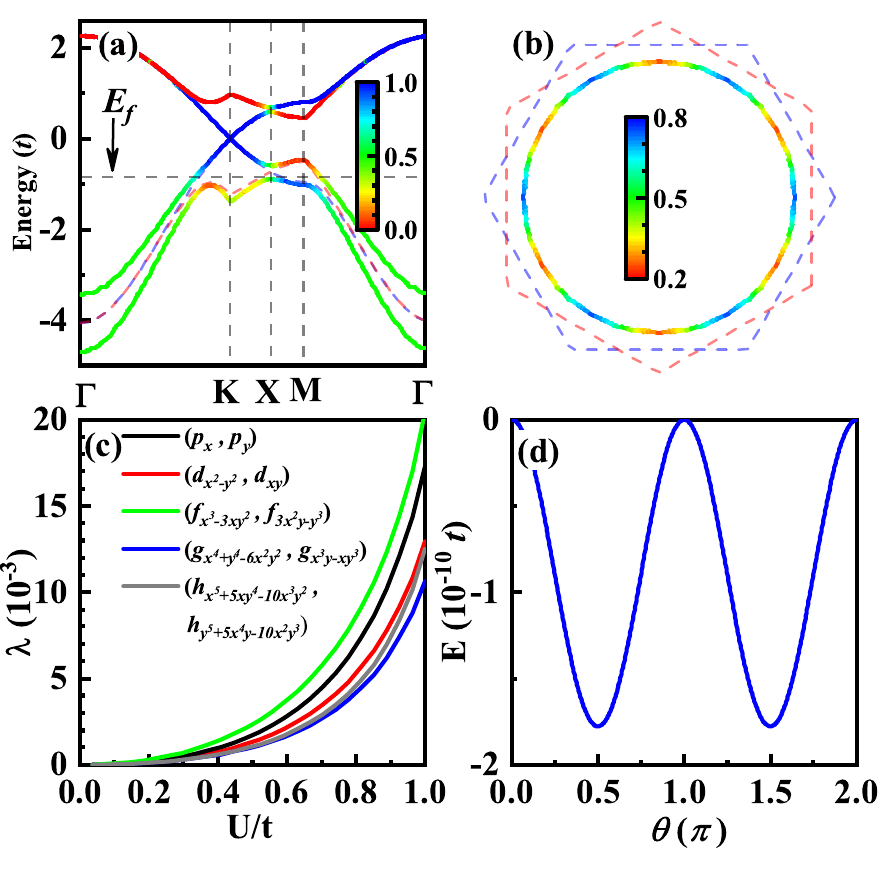}
	\caption{(Color online)  Results for the $30 \degree$-twisted bilayer honeycomb lattice. (a) Band structure along the high-symmetry lines: solid (dashed) lines for the coupled bilayer (two uncoupled monolayers). (b) FSs for $\delta=0.17$ hole doping. The colors in (a) and (b) represent layer component. (c) The $\lambda \sim \frac{U}{t}$ relations for various leading pairing symmetries at $\delta=0.17$ hole doping, (d) Mixing-phase-angle $\theta$ dependence of the energy for the degenerate $f$-wave pairings for $U=t$. }\label{bilayer}
\end{figure}

The obtained band structure for the TB-QC is shown in Fig.~\ref{bilayer}(a), in comparison with the two uncoupled band structures from the two separate monlayers, and the FS for the $\delta=0.17$ hole doping is shown in Fig.~\ref{bilayer}(b). The most prominent feature at the low hole-doping regime lies in that the two uncoupled bands from the two separate layers cross at the $X$ point (or more generally on the $\Gamma$-$X$ line) and strongly hybridize there, after which the two bands are split into the lower band and the higher band. For the hole doping level $\delta=0.17$ studied here, only the higher band crosses the Fermi level, leading to a dodecagonal-symmetric FS, as shown in Fig.~\ref{bilayer}(b).

When the Hubbard interaction is considered, we adopt the standard multi-orbital RPA approach by using the above perturbation-corrected band structure. Considering only intraband pairing between opposite momenta, we get the effective BCS Hamiltonian. Under the mean-field (MF) treatment, we can obtain the following linearized gap equation near the superconducting $T_c$ \cite{RPA1},
\begin{equation}\label{linear_eq}
-\frac{1}{(2\pi)^2}\sum_{\nu\beta}\oint dq_{\parallel}
\frac{V^{\mu\nu}_{\alpha\beta}(\mathbf{k},\mathbf{q})}
{v^{\nu\beta}_F(\mathbf{q})}\Delta_{\nu\beta}(\mathbf{q})
=\lambda\Delta_{\mu\alpha}(\mathbf{k}).
\end{equation}
where $V^{\mu\nu}_{\alpha\beta}(\mathbf{k},\mathbf{q})$ is the effective pairing interaction given in Ref. \cite{Yu_Bo2023}. This equation is solved to yield the largest pairing eigenvalue $\lambda$ and corresponding eigenvector $\Delta_{\mu\alpha}(\mathbf{k})$. The former and latter determine the $T_c$ and the gap function respectively.

The $U/t$-dependence of the largest pairing eigenvalue $\lambda$ for various pairing symmetries are shown in Fig.~\ref{bilayer}(c). The doping level is fixed at $\delta=0.17$ hole doping, at which the $f$-wave pairing is the leading pairing symmetry for the monolayer. Due to the classification according to the IRRPs of the $D_{6d}$ point group, there can be non-degenerate $s$-, $i$-, $i^{\prime}$-, $i*i^{\prime}$- wave pairing symmetries and degenerate $\left(p_x,p_y\right)$-, $\left(d_{x^2-y^2},d_{xy}\right)$-, $\left(f_{x^3-3xy^2},f_{3x^2y-y^3}\right)$-, $\left(g_{x^4+y^4-6x^2y^2},g_{x^3y-xy^3}\right)$- and $\left(h_{x^5-10x^3y^2+5xy^4},h_{5x^4y-10x^2y^3+y^5}\right)$- wave pairing symmetries for this TB-QC. Here we only show the several leading pairing symmetries.

\begin{figure}[htbp]
	\centering
	\includegraphics[width=0.5\textwidth]{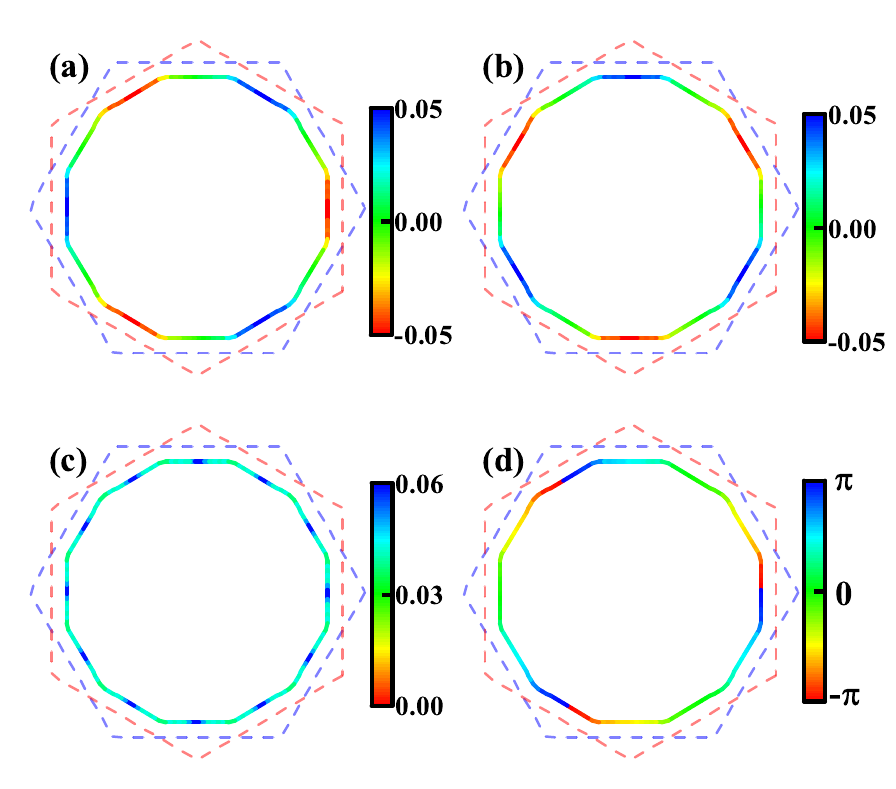}
	\caption{(Color online) Distributions of the obtained pairing gap functions on the FS of the 30\degree-twisted bilayer honeycomb lattice. The distributions of the gap functions of the obtained $f_{x^3-3xy^2}$- and $f_{3x^2y-y^3}$-wave pairings are shown in (a) and (b) respectively. The distributions of the amplitude and the phase of the gap function of the $f+if$-wave pairing state obtained are shown in (c) and (d) respectively. The hole doping level is $\delta = 0.17$ and the interaction parameter is $U=t$.}\label{fwave}
\end{figure}

From Fig.~\ref{bilayer}(c), the degenerate $\left(f_{x^3-3xy^2},f_{3x^2y-y^3}\right)$-wave pairing symmetry is the leading pairing symmetry for all the $U/t$ values shown. Note that although the leading pairing symmetry for the monolayer system and the 30\degree-twisted bilayer one is both the $f$-wave carrying pairing angular momentum 3, it is non-degenerate for the former case and degenerate for the latter case. The reason lies in that the point group has been enlarged from $D_6$ for the former case to $D_{6d}$ (isomorphic to $D_{12}$) for the latter case, and the $f$-wave belongs to the $B_2$ and $E_6$ IRRPs for the two cases respectively. The distributions of the pairing gap functions for the degenerate $f_{x^3-3xy^2}$- and $f_{3x^2y-y^3}$-wave pairing symmetries are shown on the FSs in Fig.~\ref{fwave}(a) and (b), respectively. While both $f$-wave pairing gap functions change sign with every 60\degree-rotation, they possess different symmetric and antisymmetric axes, as well as different nodal lines. Clearly, the two gap functions are mutually related by 30\degree-rotation.

The two degenerate pairing components of the $f$-wave pairing possess the same $T_c$, and would be mixed below $T_c$. We mix them as $1: e^{i\theta}$ to minimize the ground-state energy. Here we have set $U=t$ instead of $U=0.3t$ so that the condensation energy is obviously larger than the machine accuracy. Consequently, the $E(\theta)$ function shown in Fig.~\ref{bilayer}(d) is minimized at $\theta=\pm\pi/2$, leading to the $f_{x^3-3xy^2}\pm if_{3x^2y-y^3}$ ($f+if$ for abbreviation) -wave pairing state, consistent with the G-L theory. The distributions of the amplitude and phase of the gap function for the obtained $f+if$-wave pairing state are shown in Fig.~\ref{fwave}(c) and (d), respectively.

\begin{figure}[htbp]
	\centering
	\includegraphics[width=0.5\textwidth]{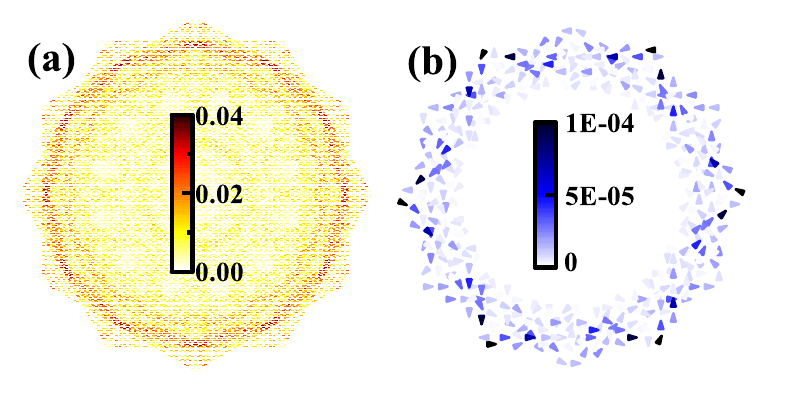}
	\caption{(Color online) Topological properties of the obtained $f+if$-wave chiral TSC on the 30\degree-twisted bilayer honeycomb lattice. (a) The real-space distribution of the squared amplitude of the wave function of a typical Majorana zero-energy state. (b) The real-space distribution of the spontaneous super current. The doping level and the interaction parameter are the same as those in Fig.~\ref{fwave}. }\label{topo}
\end{figure}

\subsection{Topological Properties}
The revised perturbational-band theory based microscopic framework adopted here possesses clear advantages over the real-space approaches in the study of the topological properties of TSCs in the TB-QC. In the weak-pairing limit which applies to most superconductors including the TB-QCs studied here, the Chern number for a fully-gapped pairing state is determined by the winding number of the pairing phase around the FS~\cite{Qi2010, Alicea2012}. As shown in Fig.~\ref{fwave}(c), the gap function of the $f+if$-wave pairing state is fully gapped, which provides condition for the realization of TSC. Further more, Fig.~\ref{fwave}(d) shows that the distribution of the pairing phase repeats three times for each run around the FS, leading to the winding number 3. Consequently, the Chern number is 3. Therefore, we have obtained here the chiral $f+if$-wave TSC with nontrivial high Chern number 3.

Usually, a chiral TSC is accompanied with Majorana chiral Fermion states and spontaneous chiral super current on the boundary. To study such topological properties, we take an open boundary condition which respects the dodecagonal symmetry of the TB-QC, as shown in Fig.~\ref{topo}. Then, setting the obtained $f+if$-wave pairing gap function as the input of the BCS MF Hamiltonian, we diagonalize the Hamiltonian and obtain the spectrum and wave functions for the Bogoliubov quasi-particles. Consequently, the real-space distribution of the squared amplitude of the wave function of a typical zero-energy Bogoliubov quasi-particle is shown in Fig.~\ref{topo} (a). Obviously, the Majorana chiral Fermion mainly distributes on the boundary of the TB-QC. Meanwhile, we calculate the real-space distribution of the current. Consequently, there exists spontaneous chiral super current which propagates along the boundary of the TB-QC, as shown in Fig.~\ref{topo} (b).

\section{Discussion and Conclusion}

In conclusion, we propose a general scheme to realize chiral TSCs from non-topological SCs through the ``{\it twistronics}''. Briefly, taking an n-fold symmetric monolayer which hosts non-topological SC with pairing angular momentum $L=n/2$, the corresponding TB-QC would host chiral TSC with the same pairing angular momentum. Besides the known example of acquiring $d+id$ TSC in the 45\degree-twisted bilayer cuprates, here we propose another example in which we can get the $f+if$- TSC in a TB-QC made of two $D_6$ symmetric monolayers which carry the $f$-wave pairing.

Note that the TB parameters adopted in our model are the same as those for the 30\degree-twisted bilayer graphene. However, the largest $U/t$ adopted in Fig.~\ref{bilayer}(c) is only about $1$, less than the real value of about $2\sim3$, because the realistic $U/t$ has gone beyond the range that can be treated in the RPA approach. In Ref. \cite{Kiesel2012}, the functional renormalization group based study reveal that the $f$-wave pairing can indeed takes place in the low hole-doped graphene with realistic interaction parameters. Then from our universal G-L theory, the $f+if$-wave chiral TSC would indeed be realized in the low hole-doped 30\degree-twisted bilayer graphene.

\section*{Acknowledgements}
This work is supported by the National Natural Science Foundation of China under the Grant Nos.12074031, 12234016, 11674025, 12074037, and 12141402. W.-Q. Chen is supported by the Science, Technology and Innovation Commission of Shenzhen Municipality (No.ZDSYS20190902092905285), Guangdong Basic and Applied Basic Research Foundation under Grant No.2020B1515120100 and Center for Computational Science and Engineering of Southern University of Science and Technology, and the National Key R and D Program of China (Grants No. 2022YFA1403700).

\section*{Note}
During the preparation of this manuscript, we became aware of a recent preprint\cite{Zhou2022} which suggested the possibility of chiral $f+if$ pairing in the context of maximally twisted double-layer spin-triplet valley-singlet superconductors.

\end{document}